\documentclass[aps,prl,twocolumn,showpacs]{revtex4}
\usepackage{graphicx}

\begin{document}
\title{From phase- to amplitude-fluctuation driven 
superconductivity in systems with precursor pairing}
\author{J.~Ranninger and L.~Tripodi}
\affiliation{ 
Centre de Recherches sur les Tr\`es Basses Temp\'eratures
 associ\'e \`a l'Universit\'e Joseph Fourier,
 C.N.R.S., BP 166, 38042 Grenoble-C\'edex 9, France}

\begin{abstract}
The change-over from phase- to amplitude-fluctuation driven 
superconductivity is examined for a composite system of free 
electrons (Fermions with concentration $n_F$) and localized electron-pairs 
(hard-core Bosons with concentration $n_B$) as a function of doping -  
changing the total concentration of charge carriers ($n_{tot}=n_F + 2n_B$).
The coupling together of these 
two subsystems via a charge exchange term induces electron pairing and 
ultimately superconductivity in the Fermionic subsystem. The difference 
in statistics of the two species of charge carriers has important 
consequences on the doping mechanism, showing an onset temperature $T^*$ 
of incoherent electron pairing in the Fermionic subsystem 
(manifest in form of a pseudogap), which steadily decreases with decreasing 
$n_{tot}$. Below $T^*$ this electron pairing leads, in the normal phase, 
to electron-pair resonant states (Cooperons) with quasi-particle 
features which strongly depend on $n_{tot}$. For high concentrations,  
where $n_B \simeq 0.5$, correlation effects between the hard-core Bosons 
lead to itinerant Cooperons having a heavy mass $m_p$, but are long-lived. 
Upon reducing the concentration of charge carriers and consequently $n_B$, 
the mass as well as the lifetime of those Cooperons is considerably reduced. 
As a result, for high values of $n_B$, a superconducting state below $T^*$ 
sets in at a $T_c$, being controlled by the phase stiffness 
$D_{\phi}=\hbar^2 n_p/m_p$ of those Cooperons, where $n_p$ denotes their 
density. Upon reducing $n_{tot}$, the phase stiffness steadily increases, 
and eventually exceeds the pairing energy $k_B T^*$. There, the Cooperons 
loose their well defined itinerant quasi-particle features and 
superconductivity gets controlled by amplitude fluctuations. The resulting
phase diagram with doping is reminiscent of that of the phase 
fluctuation scenario for high $T_c$ superconductivity, except that in our 
scenario the determinant factors are the mass and the lifetime of the 
Cooperons rather than their density. 

\end{abstract}
\maketitle

\section{I Introduction}

The features which characterize classical low temperature 
superconductors are the disappearance, above a certain critical temperature 
$T_c$, of a gap in the density of states (DOS) of the electrons,  occurring 
simultaneously with the disappearance of the magnetic field expulsion (the 
Meissner effect) and standard Fermi liquid behavior in the normal phase 
above $T_c$. A further characteristic is the practical impossibility 
to change significantly the value of $T_c$ upon changing the concentration 
of charge carriers, because of $T_c$ being largely determined by the density 
of states (DOS) at the Fermi level, which generally is not expected to 
change much with doping.

None of these features are observed in the high temperature superconductors 
(HTSC). The opening of a gap in the DOS occurs gradually, as a multitude 
of different experiments\cite{Timusk-99}) show. This gap initially emerges 
in form of a pseudogap -  a dip in the DOS - below a certain temperature 
$T^*$, which, depending on doping, can be much above $T_c$. Upon lowering 
the temperature and approaching $T_c$, this pseudogap smoothly joins the 
superconducting gap, as is evident from its angular variation near the Fermi 
surface. If the opening of the pseudogap and of the superconducting gap  
represent  different physical manifestations of one and the same pairing 
mechanism, an interplay between these manifestations of electron 
pairing in the two phases is to be expected and is in fact observed in form 
of remnant effects, such as: 

(i) A remnant of magnetic field expulsion is seen in form of a transient 
Meissner effect  several tens of degrees above $T_c$, judging from 
the optical conductivity in the Tera Hertz regime\cite{Corson-99}. This 
points toward long-lived diamagnetic fluctuations, which have been attributed 
to the presence of long-lived diffusing vortices above $T_c$\cite{Xu-00}, 
as well as to phase uncorrelated diamagnetic regions which act as precursors 
to the true Meissner state below $T_c$\cite{Iguchi-01}. Experiments, invoking 
Andreev reflections to interpret the enhanced tunneling conductance in the 
pseudogap phase\cite{Deutscher-99,Choi-00}, point toward phase 
uncorrelated pairing above $T_c$. 

(ii) Remnants of local electron-pairing in the c-axes optical response 
(orthogonal to the $CuO_2$ planes)\cite{Homes-93} in the pseudogap phase,
are seen in the superconducting phase. Similarly, the c-axes component 
of the electronic kinetic energy 
is getting reduced upon entering the superconducting phase\cite{Basov-01}, 
provided the normal state exhibits pseudo gap features. The doping dependence 
of the c-axis penetration depth, being qualitatively similar to that of the 
basal plane, suggests that both are strongly influenced by the pseudogap 
features of the normal state\cite{Panagopoulos-00}. Tunneling measurements 
have indicated remnants of the pseudogap which continue to coexist with 
the superconducting gap in the superconducting phase. The latter disappears 
at $T_c$, while the former remains\cite{Krasnov-00}. 

The question whether such findings favor or not a common origin of 
pseudogap and of the superconducting gap is presently still being 
debated\cite{Suzuki-00,Fischer-01}. 

The features in the HTSC, involving the interplay of the pseudogap phase 
and of the superconducting phase, are highly doping dependent. $T^*$ 
steadily decreases with increased hole doping, while $T_c$ shows an equally
steady rise until the two approach each other. $T_c$ then bends over and 
follows the descent of $T^*$ upon further doping and approaching the 
optimal/overdoped regime. $T_c$ plotted as a 
function of the phase stiffness (determined by the square of the inverse 
penetration depth), involving the ratio of the density of superfluid carriers 
over their mass, shows a universal linear behavior\cite{Uemura-89}. 

The main emphasis in this paper will be to analyse the doping dependence 
of $T^*$ and $T_c$ within a precursor pairing scenario. Very little about 
that doping dependence is known for such a scenario when based on single 
component systems, such as the negative U Hubbard model or the effective 
BCS Hamiltonians, extended to strong coupling. In such studies, doping is 
frequently introduced ad hoc, by assuming a doping dependent electron 
hopping or inter-electron attraction, in view of simulating a physics close 
to a Mott transition. In the present paper we shall examine such doping 
dependent effects without making any such ad hoc assumptions. As we shall 
see, considering a mixture of itinerant electrons (Fermions) and localized 
electron-pairs (hard-core Bosons) coupled together via a charge exchange 
term, is capable of reproducing the doping dependent features of $T^*$ and 
$T_c$ specified above. The essential new features introduced in this 
two-component scenario are the difference in statistics of the two 
components and the hard core features of the short range two-electron 
resonances (Cooperons) in the itinerant electron subsystem which result 
as a consequence of the charge exchange coupling.
The variation of $T_c$ is dictated by correlation effects of the Cooperons.
 
This two-component scenario has  some similarity to the single-component 
scenarios with 
attractive inter-particle interaction mentioned above, in as far as it can be 
viewed as a two-fluid picture of existing preformed pairs and unpaired 
electrons\cite{Beck-01}. The electron-pair resonant states which result in 
such a scenario have certain features which are reminiscent of localized 
resonance impurity states seen in the HTSC, and which arise when $Cu$ atoms 
are substituted by non-magnetic atoms, like $Zn$\cite{Kruis-01}. Yet, the 
electron-pair resonant states which we are considering here, have the 
essential potentiality of becoming itinerant and thus to lead to a 
superconducting phase controlled by excitations of electron-pairs with 
finite momenta rather than pair breaking. 

In section II we discuss the interplay between phase and amplitude 
fluctuations in systems with precursor pairing, contained in the spectral 
properties of the Cooperon propagator. This permits us to make a connection 
with 
the phase fluctuation scenarios, which have been widely discussed in the 
literature. In section III we  briefly outline the model and the Green's 
function formalism which we adopt in order to treat the hard-core nature of 
the resonant electron-pairs. In section IV we present the results for the 
temperature variation of the density of states over a wide doping regime and 
analyse the doping variation of $T^*$. In section V 
we explore the spectral properties of the two-particle excitations 
and compare the low and high density regime of the Cooperons as far 
as their quasi-particle properties are concerned. In section VI we discuss 
the thermodynamics of the pseudogap phenomenon in terms of the specific heat
and entropy for different doping regimes. Finally in section VII 
we give a summary of our findings which are characteristic of precursor 
pairing systems, involving Fermionic as well as Bosonic charge carriers.

\section{II The Cooperon within a phase fluctuation scenario}

Based on the different experimental results mentioned in the Introduction,
which clearly indicate that the  physics of those HTSC is not BCS like (over 
at least a wide regime of doping), it is widely agreed 
upon that the onset of the superconducting state in the underdoped regime 
in those HTSC should be controlled by phase rather than amplitude 
fluctuations\cite{Emery-95,Chakraverty-99,Chakraverty-00}. This supposes 
the existence of 
local superconducting droplets with a given phase, pre-existing above $T_c$ 
and evolving into a macroscopic phase locked state upon entering the 
superconducting state. Such a situation can be realized provided that the 
fluctuations of the phase of the macroscopic superconducting wave function are 
less costly in energy than the fluctuations of the amplitude of the 
electron-pairs, which describe pair breaking. To within a first 
approximation, this so-called  phase fluctuation scenario 
is generally described within a hydrodynamic formulation of a spatially 
fluctuating  phase $\phi$ and its conjugate variable $\delta n_s$, which 
describes the spatial fluctuations of the superfluid particle density $n_s$. 
The corresponding effective Hamiltonian is given by
\begin{equation}
H=\frac{1}{2}D_{\phi} (\nabla \phi)^2 + \frac{1}{2\chi n_s^2} (\delta n_s)^2,
\end{equation} 
where $D_{\phi}=\hbar^2 (n_s/m_s)$ denotes the superfluid phase stiffness, 
$\chi$ the compressibility, and $m_s$ their respective mass. The temperature 
which controls the phase order of such a system is given by 
$k_B T_{\phi} \simeq D_{\phi} a$, with $a$ being either given by the 
coherence length $\xi$ for 3D systems or by the interlayer distance in 
layered compounds such as HTSC. 

This phase fluctuation scenario is frequently discussed in conjunction with 
the so called 
BCS-BEC cross-over phenomenon\cite{Randeria-95}, where, as a function of 
the strength of the inter-particle attraction, one passes from a BCS state  at 
weak attraction to a Bose Einstein condensation (BEC) of tightly bound 
electron-pairs in the limit of strong attraction. The physics for that has 
been widely studied on the basis of effective BCS and  negative U Hubbard 
Hamiltonians, aiming to treat the single-particle and two-particle features 
on the same footing. The pseudogap in such a scenario arises from short range 
electron-pairing, correlated over a finite time scale, comparable 
to the energy scale of the zero temperature superconducting gap. According 
to a general theorem (due to Bogoliubov) and based on the singular behavior 
of the occupation number of electron-pairs with small momenta which 
signal bound states, such electron pairing ought to  survive below 
$T_c$\cite{Cherny-99}. A possible experimental verification for that might 
come from the socalled {\it {peak-dip-hump}} feature in 
ARPES\cite{Norman-97}, which shows a spectral behavior upon entering the 
superconducting phase, where a sharp peak (related to superconducting 
correlations of the low energy excitations) emerges out of the broad 
incoherent background, characterized by a broad hump and representing 
remnants of the pseudogap phase.

An attempt to formulate the problem of local pairing as a prerequisite of 
superconductivity was made a long time ago by generalizing the mean field BCS 
formalism such as to cover the regime above as well as below 
$T_c$\cite{Kadanoff-61}. Instead of using the order parameter, the propagator 
for the electron-pairs, also called Cooperons, is introduced and treated  
on the same level as the single electron propagator. Within such a formalism, 
and on a quite general basis, the pseudogap phase of the HTSC has been 
examined\cite{Tcherni-97}, invoking the mutual feedback effect between 
the single- and the two-particle properties via the introduction of some 
effective gap above $T_c$. 

A different procedure was followed by proposing a specific structure of 
the Cooperon propagator\cite{Giovannini-01} of the form
\begin{equation}
C({\bf r}, t)= C_a exp(-r/r_0) + C_{\phi}\langle e^{i\phi({\bf r},t)} 
e^{-\phi(0,0)}\rangle,
\end{equation}
separating amplitude from phase correlations in an additive way. The first 
term, representing a rather rapidly decreasing function with $r$, describes
local pair amplitudes which are treated in a time independent fashion. 
The second term describes the 
phase correlations, which, in the low frequency limit, are approximately 
expressed by $C_{\phi} exp(-r/\xi(T))$ where  $\xi(T)$ denotes the temperature 
dependent coherence length. Attributing that latter contribution of the 
Cooperon propagator to a $2D$ $XY$ physics above the Kosterlitz-Thouless 
critical temperature $T_{KT}$, establishes a link with the phase fluctuation 
scenario. The doping behavior of such precursor systems as the HTSC is then 
monitored by parameterizing 
the relative weight of the coefficients $C_a, C_{\phi}$ together with the 
relative spatial extent of the two contributions of $C({\bf r}, t)$ such 
that it describes:

(i) an underdoped regime, characterized by primarily phase fluctuation 
controlled onset of superconductivity with a large temperature regime for 
the pseudogap phase, 

(ii) an optimal/overdoped  regime, characterized by a gradual 
disappearance, upon increased doping, of the pseudogap phase and a 
superconducting phase, controlled by amplitude correlations. 

Considering the origin of the pseudogap as being due to pair fluctuations, 
also called precursor pairing (not to be confused with preformed pairs), the 
characteristic temperature $T^*$ below which the opening of the pseudogap 
occurs for such single-component scenarios with inter-particle attraction, 
scales with the strength of that 
interaction\cite{Micnas-95,Levin-01,Metzner-01,Kopec-02}. However, as far as
the concentration dependence of $T_c$ and $T^*$ is concerned within such 
scenarios, it invariably shows that both $T^*$ and $T_c$ follow the same 
trend\cite{Tremblay-99,Tremblay-01} upon varying the number of charge 
carriers. This is clearly the opposite to what is found in the HTSC. 
Potential dimensionality changes, linked to the change-over from 
underdoped to overdoped materials, can not remedy this situation either, 
since as far as $T^*$ is concerned, it is determined by essentially local 
physics and thus independent on any dimensional aspects. As far as $T_c$ 
is concerned, even upon assuming a cross-over, with reduced doping, to a 
Kosterlitz-Thouless critical temperature behavior in the underdoped regime,  
$T_{KT}$ would still follow the same doping trend as $T^*$.
It thus seems likely that correlation effects are indispensable in 
determining the doping dependence of $T_c$ versus $T^*$ in such precursor 
scenarios. 

\section{III The model and the technique employed} 

The scenario of a mixture of itinerant Fermions (band electrons) and localized 
hard-core Bosons (bound electron-pairs) will be described on the basis of the 
so-called Boson-Fermion model (BFM). This model presents a paradigm for 
interacting electron systems where two-particle resonant states are 
expected to occur due to the interaction of the electrons with certain Bosonic 
modes and where such two-particle resonant states act as precursor to a 
transition into a superconducting state. The underlying 
physics behind this model\cite{Ranninger-02}, as it was initially 
conceived\cite{Ranninger-85}, is that of electrons strongly coupled to local 
phonons, which act as such Bosonic modes. This results in self-trapped 
entities, comprising the charge carriers and the surrounding clouds of Bosonic 
excitations, in form of resonant pair states inside a system of itinerant 
electrons. Such a BFM scenario is not in any way restricted to electronic 
systems undergoing a superconducting transition but ought equally well 
apply to electron-hole pairing in semi-conductors\cite{Mysyrowicz-96}, and 
low density nuclear matter with isospin singlet pairing\cite{Schnell-99}.
Moreover, it was in a similar spirit that such a Boson-Fermion mixture 
scenario has been derived recently for (i) the Hubbard model with intermediate 
repulsive coupling\cite{Auerbach-02}, (ii) the exchange interaction between 
spinon singlets of resonating valence bond (RVB) electron-pairs and 
holons\cite{Kochetov-02} and (ii) for 
entangled atoms in squeezed states in molecular Bose Einstein condensates in 
traps\cite{Yurovski-02}. More generally, the BFM has been employed in 
attempts to bosonize an intrinsically Fermionic system\cite{Lee-94}.

The model Hamiltonian describing the Boson Fermion scenario is given by 
\begin{eqnarray}
H_{0} & = & (D-\mu )\sum _{i\sigma }c_{i\sigma }^{+}c_{i\sigma } 
 + (\Delta _{B}-2\mu )\sum _{i}(\rho_{i}^{z}+\frac{1}{2})\nonumber\\
& + & t\sum _{i\neq j,\: \sigma }c_{i\sigma }^{+}c_{j\sigma }
+ v\sum _{i}\left( \rho^{+}_{i}c_{i\downarrow }c_{i\uparrow }+
\rho^{-}_{i}c_{i\uparrow }^{+}c_{i\downarrow }^{+}\right) 
\end{eqnarray}
where the localized hard-core Bosons are represented by pseudo spin-1/2 
operators $[\rho_i^+,\rho_i^-,\rho_i^z]$ and the itinerant electrons by 
$[c_{i\sigma},c^+_{i\sigma}]$. $v$ denotes the strength of the 
onsite hybridization between the two types of charge carriers and  $t$ the 
hopping integral for the itinerant electrons with a band half width $D$. The 
full band width $2D$ will be used as the energy unit through out this paper. 
The energy level of the localized Bosons is given by $\Delta_B$ and the 
chemical potential $\mu$ is chosen to be common to both species, such as to 
ensure total charge conservation, $n_{tot}=n_F+2n_B$. $n_F$ and $n_B$ denote 
the occupation number per site of the electrons (including up and down spin 
states) and of the hard core-Bosons. 

The opening of the pseudogap and its temperature dependence on 
the basis of this  model, for a fixed concentration and upon 
neglecting the hard-core nature of the Bosons has been studied 
previously\cite{Ranninger-95a,Devillard-00,Domanski-01}. In a study, 
based on the dynamical mean field approach\cite{Robin-98}, the hard-core 
nature of the Bosons could be taken into account. But then, the itinerancy 
of the Bosons could not be treated within such a scheme which restricted 
this study to purely amplitude fluctuations driven superconductivity.             
In the present work we shall account for both, the hard-core nature of 
the Bosons as well as their potentiality becoming itinerant and shall 
study the pseudogap characteristics as a function of total carrier 
concentration. The present study follows closely the previous  
self-consistent diagrammatic 
approach\cite{Ranninger-95a}, but  generalizes it such as to take 
into account the hard-core nature of the Bosons. We adopt for that 
purpose the diagrammatic technique which had been developed for spin 
systems\cite{Vaks-68,Izyumov-88} (with their convention 
$[\rho^+,\rho^-]=\rho^z$) and for which it was shown that the 
usual Wick theorem had to be generalized  to 
\begin{eqnarray}
& &\left\langle T\left\{ \rho_{1}^{\alpha _{1}}(\tau _{1})\ldots 
\rho^{-}_{0}(\tau)
 \ldots \rho_{n}^{\alpha _{n}}(\tau _{n})\right\} \right\rangle _{0}\nonumber \\ 
& = & K_{01}^{0}(\tau -\tau _{1})
\left\langle T\left\{ \left[ \rho_{1}^{\alpha _{1}},\rho_{0}^{-}
\right] _{\tau _{1}}\rho_{2}^{\alpha _{2}}(\tau _{2})
\ldots \rho_{n}^{\alpha _{n}}(\tau _{n})\right\} \right\rangle _{0}\nonumber \\
 & + & K^{0}_{02}(\tau -\tau _{2})
\left\langle T\left\{ \rho_{1}^{\alpha _{1}}(\tau _{1})\left[ \rho^{\alpha _{2}}_{2},\rho_{0}^{-}\right] _{\tau _{2}}\ldots \rho_{n}^{\alpha _{n}}(\tau _{n})
\right\} \right\rangle _{0}\nonumber \\
  & + &  \ldots 
\end{eqnarray}
with
\begin{eqnarray}
K_{11'}(\tau -\tau ') & = & \delta _{11'}K^{0}(\tau -\tau ')\nonumber \\
K^{0}(\tau -\tau ') & = & \frac{\left\langle T\rho^{-}(\tau )\rho^{+}(\tau ')\right\rangle _{0}}{\left\langle \rho^{z}\right\rangle _{0}}=\nonumber\\
 & = & e^{-E_{0}(\tau -\tau ')}n(x_{0})\theta (\tau -\tau ')+\nonumber\\
 & + & e^{-E_{0}(\tau -\tau ')}(1+n(x_{0}))\theta (-\tau +\tau ')\nonumber\\
n(x_{0})& = &  \frac{1}{e^{x_{0}}-1}, \;\;\;x_{0}  =  -\beta E_{0}.
\end{eqnarray}

Recursively applying this procedure of the modified Wick theorem, the 
remaining 
multi-spin correlation functions are transformed step by step into a sum 
of products of $K_{ij}^0$ multiplied with thermal averages of the type 
$\langle \rho_i^z....\rho_n^z\rangle_0$ (evaluated with respect to the 
unperturbed Hamiltonian: $(\Delta_B-2\mu)\sum_i(\rho_i^z+\frac{1}{2 })$). 
These thermal averages are expressed in terms of a set of cumulants, 
the first few of which are given by:
\begin{eqnarray}
\langle\rho^z_1(\tau)\rangle &=& b =-\frac{1}{2} + n_B \nonumber 
\qquad\qquad  \\
\langle \rho^z_1(\tau)\rho^z_2(\tau) \rangle &=& b^2 + b'\delta_{1,2},\quad 
b^2 + b'=\frac{1}{4} \nonumber \\
\langle \rho^z_1(\tau)\rho^z_2(\tau)\rho^z_3(\tau) \rangle&=&
b^3 + b b'(\delta_{1,2}+\delta_{2,3}+\delta_{3,1}) \nonumber \\
 + b''\delta_{1,2}\delta_{2,3}\; &,& 
 \;  b^3 + 3bb' + b''= \frac{1}{4}b
\end{eqnarray}
\begin{figure}
\begin{center}
\includegraphics[width=3.0in,height=3.0in]{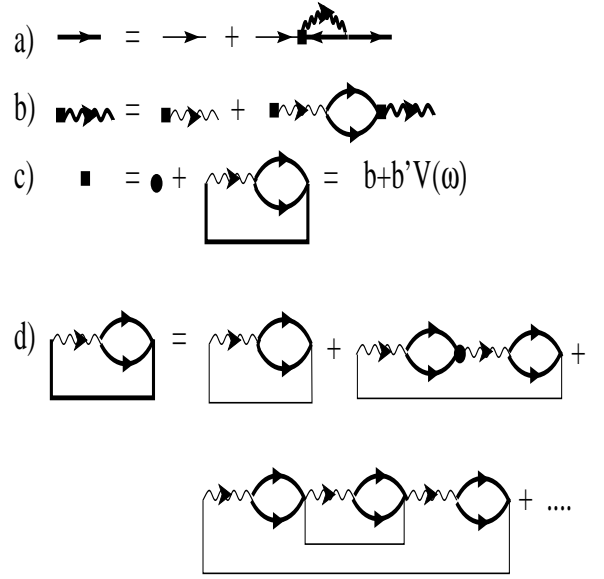}
\end{center}
\caption{\label{Diagrams} Diagrammatic representation of the Fermion (a), 
 Boson (b) and the vertex correlation functions (c+d).}
\end{figure}
Keeping only the first two cumulants $b$ and $b'$ gives rise to the set of 
diagrams, illustrated in Fig. 1, and describe the set of self-consistent 
equations determining the Fermion (Fig. 1a) and Boson Green's functions 
(Fig. 1b) $G(k,\omega _{n})$ and $K(q,\omega _{m})$. The vertex depicted by 
the full square in those figures is made 
up of two contributions: one arising from the cumulant $b$ and depicted  
by a full circle and one arising from the cumulant $b'$, illustrated 
by the second contribution to this vertex. Both those vertex contributions 
have to be determined selfconsistently. For the contribution arising from the 
cumulant $b$ it simply is $b=-1/2 +n_B$. For the contribution arising from 
the second cumulant, given by $b' V(\omega)$ the selfconsistent equation 
corresponding to the set of diagrams illustrated in Fig. 1d has to be solved.
To within this approximation of cumulants, this leads to the following set of 
equations:

\begin{eqnarray}
G(k,\omega _{n})&=&\frac{1}{i\omega _{n}-
\varepsilon _{k}-\Sigma (k,\omega _{n})},\nonumber 
 \\
K(q,\omega _{m})&=&\frac{b+b'V(\omega_m)}{i\omega _{m}-E_{0}-
(b+b'V(\omega_m))\Pi (q,\omega _{m})},\nonumber \\
G^{0}(k,\omega_n)&=&\frac{1}{i\omega _{n}-\varepsilon _{k}};
\;\; \; K^{0}(\omega_m)=\frac{b}{i\omega _{m}-E_{0}},\nonumber \\
V(\omega_m)&=&V_0(\omega_m) + K_0(\omega_m)\frac{1}{N}\sum_q
\Pi^2(q,\omega_m)K(q,\omega_m),\nonumber\\ 
V_0(\omega_m)&=&\frac{1}{N}\sum_q \frac{\Pi (q,\omega _{m})}
{i\omega _{m}-E_{0}}
\label{GFEq1}
\end{eqnarray}
with $E_0=\Delta_B-2\mu$ and the bare electron dispersion 
$\varepsilon_{\bf k}=D(1-\frac{1}{Nz}\sum_{\langle r_i \neq r_j \rangle}
e^{i{\bf {k(r_i-r_j)}}})-\mu$. The self energies for the Fermions and 
hard-core Bosons are given by:
\begin{eqnarray}
\Sigma ({\bf k},\omega _{n})&=& -\frac{v^{2}}{\beta N }
\sum _{{\bf q},\omega _{m}}G({\bf q-k},\omega _{m}-\omega _{n})
K({\bf q},\omega _{m})\nonumber \\
\Pi ({\bf q},\omega _{m})&=& \frac{v^{2}}{\beta N}
\sum _{{\bf k},\omega _{n}}G({\bf q-k},\omega _{m}-\omega _{n})
G(k,\omega _{n})
\label{GFEq2} 
\end{eqnarray}

This set of equations represents a generalization of the usual 
self-consistent RPA equations for this 
BFM-problem\cite{Ranninger-95a,Devillard-00} when restricting oneself to
the lowest order approximation involving only the cumulant $b$. The 
contributions arising from higher order comulants bring in frequency 
dependent vertex corrections $V(\omega)$. Qualitatively, the hard-core 
nature of the 
Bosons is already contained in the lowest approximation, due to the 
appearance of the factor $b$ in the expression for the hard-core Bose Green's 
function. Given the considerable complexity in solving these equations we 
have, for a  restricted set of values for the temperature and the 
total particle concentration, compared the results which arise from the 
full set of equations (\ref{GFEq1}, \ref{GFEq2}) with those arising form 
the lowest 
order approximation, i.e., keeping only the cumulant $b$. 
The results being qualitatively the same, and given our aim to present  
only very robust qualitative features of the physics we want to discuss here, 
we report in the following exclusively the results based on this lowest order 
approximation.

The Green's functions for the Fermions and the hard-core Bosons are defined by
\begin{eqnarray}
G_{i,j}(\tau,\tau')&=& - \langle T[c_{i\sigma}(\tau)
 c^+_{j\sigma}(\tau')] \rangle \nonumber \\ 
&=&\frac{1}{N\beta }\sum _{{\bf k},n}e^{i{\bf k}({\bf {r_i-r_j}}) - 
i\omega _n(\tau -\tau ')}
G({\bf k},\omega _n), \nonumber  \\
K_{i,j}(\tau ,\tau ')&=&\left\langle T\left[ \rho^-_i(\tau )
\rho_j^+(\tau ')\right] \right\rangle \nonumber \\ 
&=& \frac{1}{N\beta }
\sum _{{\bf q},m}e^{i{\bf q}({\bf {r_i-r_j}})-i\omega _{m}(\tau -\tau ')}
K({\bf q},\omega _{m}).
\end{eqnarray}
\noindent
where $\omega_n=\pi(2n+1)/\beta $ and $\omega_m=\pi2m/\beta$ denote 
the Matsubara frequencies for Fermions and Bosons respectively, 
$n$ and $m$ running over all integers from $-\infty$ to $+\infty$. 
The expressions for the self energies for the Fermions, 
$\Sigma ({\bf k},\omega _{n})$ and for the hard-core Bosons, 
$\Pi ({\bf q},\omega _{m})$ differ from the standard ones for ordinary Bosons 
by a change in sign due to the Wick theorem for hard-core Bosons. The 
effect of the cumulants however corrects this sign change in the end 
because of the presence of the factor $b$ in the numerator of the Bose 
Green's function. Fixing $n_{tot}$, Eqs. (7) are solved numerically on 
the Matsubara axes and  the resulting Green's functions and selfenergy
functions are then analytically 
continued onto the real frequency axes, via the usual Pad\'e type procedure. 
The Green's functions $G$ and $K$ are linked to the occupation numbers 
$n^F_{\bf k} = (2/\beta)\sum_{\omega_n}
e^{-i \omega_n 0^-} G({\bf k},\omega_n)$ 
and  
$n^B_{\bf q}=  (2/\beta)\sum_{\omega_m}
e^{-i \omega_m 0^-} K({\bf q},\omega_m)$ for the Fermions and Bosons 
respectively, with $n_{F,B}=\frac{1}{N}\sum_{\bf k} n^{F,B}_{\bf k}$. 

As a typical example for the present study we choose the energy level of 
the hard-core Bosons to lie in the center of the band of itinerant 
electrons ($\Delta_B=1$)  and assume a small value of the exchange 
coupling constant ($v^2=0.02$). Requiring the chemical potential to 
lie slightly below the Bosonic level ($\mu \leq 0.5$), assures us that 
upon changing the total number of charge carriers 
$n_{tot}$ from 2 to 1, we recover a situation which, as far as the density 
of electrons is concerned, mimics the situation encountered in HTSC over 
a wide doping regime. We shall for that reason adopt the terminology, widely 
used in connection with studies on the HTSC's, and refer to the doping 
regime $2 \geq n_{tot} \geq n_0$ as the underdoped regime and  
$n_{tot} \leq n_0$ as the optimal/overdoped regime, with $n_0 \simeq 1.1$ 
for our choice of parameters. We furthermore restrict 
the momentum summations over a 1D Brillouin zone with 200 $k$-points. This 
is justified, since we are interested here in only very general features 
of the pseudogap phenomenon. If our results, have any bearing 
on the physics of HTSC, they should apply to regions in momentum space 
where the pseudogap phenomenon is most pronounced i.e., near the so-called 
{\it hot spots} around the $M$ points in the Brillouin zone of the basal 
plane. There they could describe  $d$ wave symmetry pseudogap behavior along 
one of it's lobs along a direction $[0,0] - [0,\pi]$ and their equivalents, 
traversing the $M$ points.  Our results could thus possibly be compared with 
ARPES spectra for wave vectors along such directions in $k$ space, as well 
to transport measurements along the {\underline {same}} directions.  
\begin{figure}
\begin{center}
\includegraphics[width=3.0in,height=3.5in]{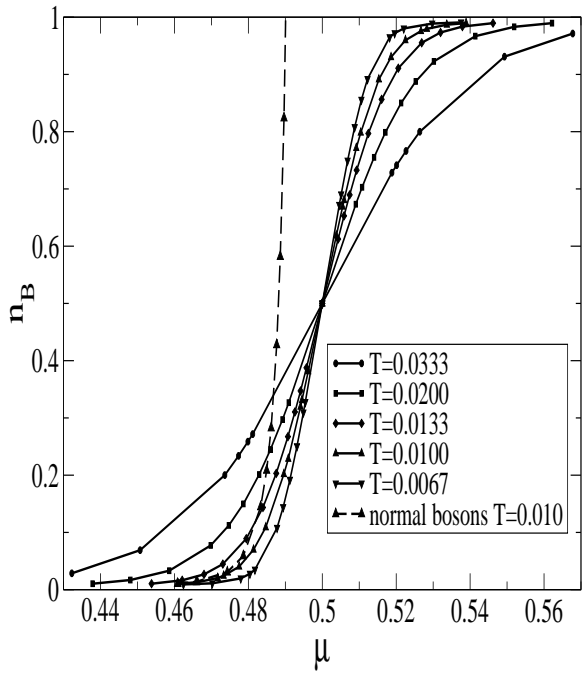}
\end{center}
\caption{\label{nBmu} Comparison of the variation of the number of normal 
to hard-core  Bosons as a function of the chemical potential for several 
temperatures.}
\end{figure}

\section{IV The doping and temperature dependence of the pseudo-gap}

We present in the following the results of the solution of the set of 
self-consistent equations (\ref{GFEq1}, \ref{GFEq2}), together with a 
self-consistent determination of the lowest cumulant $b$, Eq. (6). This 
permits us to determine the electronic 
DOS as a function of temperature for different ``doping rates'',  given by  
$n_{tot}$, or alternatively, by the depletion of the Fermi sea away from 
half filling, given by $(1-n_F)$. In this BFM scenario, doping 
influences both the number of itinerant electrons $n_F$ as well as the number 
of bound electron-pairs $n_B$. This is not an unrealistic premise as 
far as the HTSC are concerned, since it has been experimentally established 
that doping does not occur exclusively in the $CuO_2$ planes but involves 
also the dielectric layers between them. This is born out by  XPS studies 
which permit to determine the relative change with doping of 
the population of $Cu^+$ versus $Cu^{++}$ ions \cite{Tolentino-92}. 
Further indications that doping occurs in a multi-component system comes 
from measurements of the size of the Fermi surface volume\cite{Fink-02} which 
show that the universal curve for $T_c$ as a function of doping is shifted 
downwards in doping as compared to its dependence on the chemical doping 
rate. And finally, site dependent XAFS studies\cite{Nuecker-98} show 
that in order for the superconducting phase to materialize, doping must 
necessarily involve holes located outside the metallic $CuO_2$ planes. 
Further evidence for the 
existence of two species of different charge carriers, itinerant ones (giving 
rise to a Drude  peak) and localized ones (giving rise to a peak in the far 
infrared regime), comes from reflectivity measurements\cite{Calvani-00}.  

Given our choice of the Boson level falling in the middle of the  
band of itinerant electrons, fixes the Fermi level such that we have  
the situation of a half filled band for $n_{tot}= 2$. 
Upon hole doping we move the chemical potential downwards from its value 
at $D$, which introduces holes in the electronic subsystem and at the same 
time diminishes $n_B$. This variation of $n_B$ as function of $\mu$ is 
illustrated in Fig. \ref{nBmu} for different temperatures. The bound 
electron-pairs, 
being hard-core Bosons lead to a fully symmetric situation for particle and 
hole doping for this choice of parameters. We also illustrate in this figure 
the variation of $n_B$ with $\mu$ for ordinary Bosons, which significantly 
differs from that of hard-core Bosons.

\begin{figure}
\begin{center}
\includegraphics[width=3.0in,height=4.0in]{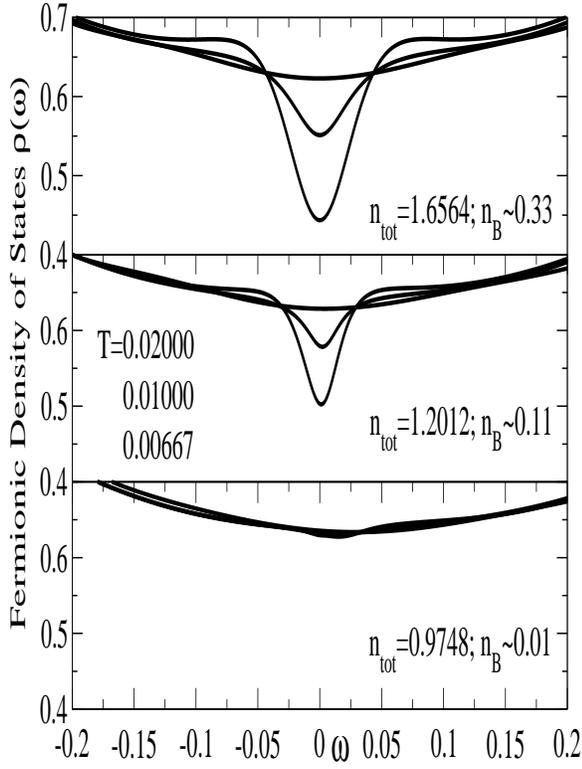}
\end{center} 
\caption{\label{rho(T)} Evolution of the pseudo-gap near the chemical 
potential as a function of temperature for $\Delta_B=1$ and different 
doping concentrations $n_{tot}$.} 
\end{figure}

\begin{figure}
\begin{center}
\includegraphics[width=3.0in,height=3.0in]{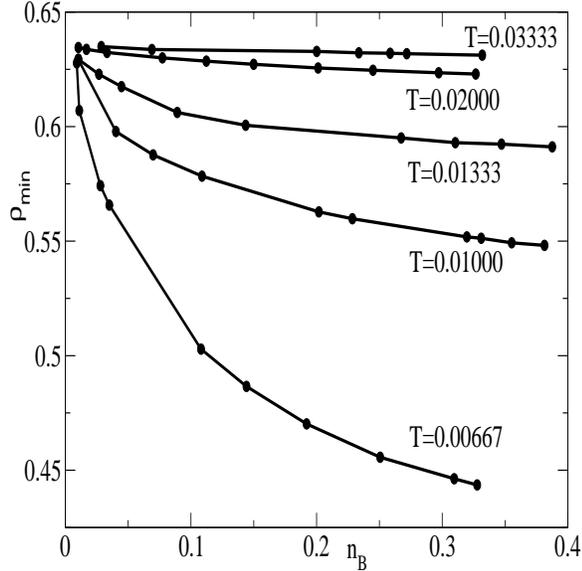}
\end{center} 
\caption{\label{rhomin(nB)}  Dependence of the minimum in the DOS on the 
concentration of 
hard-core Bosons $n_B$ for different temperatures and  $\Delta_B=1$.} 
\end{figure}

\begin{figure}
\begin{center}
\includegraphics[width=3.0in,height=4.0in]{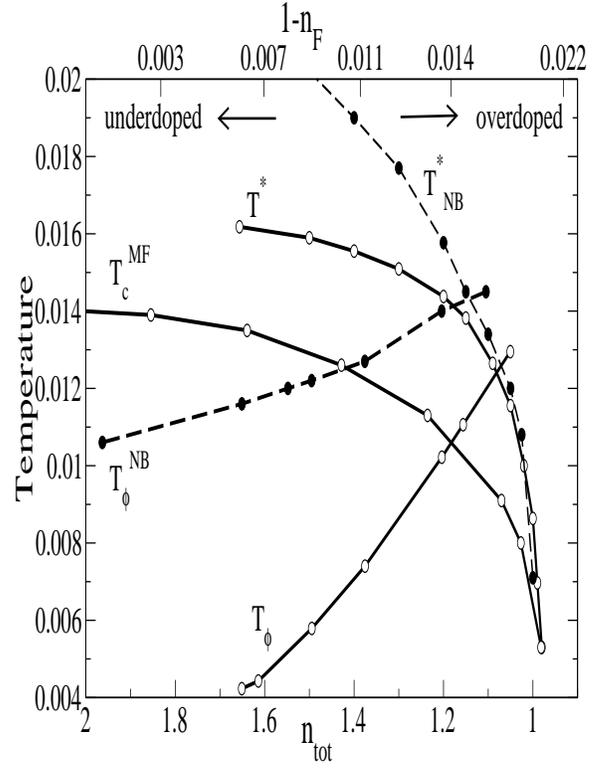}
\end{center} 
\caption{\label{TPG} Doping dependence of $T^*$, compared with the 
mean field critical temperature $T_c^{MF}$ and the ``phase fluctuation'' 
temperatures $T_{\phi}$. For comparison we also illustrate this doping 
dependence by $T^*_{NB}$ and $T_{\phi}^{NB}$  when treating the Bosons 
as normal Bosons.} 
\end{figure}

In order to study the evolution of the pseudo-gap as a function of 
temperature and doping, we evaluate the spectral function of 
the single-particle Fermionic Green's function 
$A_F({\bf k},\omega)= 2{\it Im}G({\bf k},i\omega_n 
\rightarrow \omega+i\delta))$ 
which, after integrating over all wave-vectors in the Brillouin zone gives us  
the DOS, $\rho(\omega)$. In Figs. \ref{rho(T)} we plot the evolution of the 
pseudo-gap near the chemical potential (corresponding to 
$\omega=0$) as a function of temperature and for several doping concentrations 
$n_{tot}= \simeq $ 1.66, 1.20 and 0.97.  
In order to determine $T^*$ as a function of $n_B$ and  $n_{tot}$ 
respectively, we illustrate in Fig. \ref{rhomin(nB)} the minimum of the 
dip in the DOS 
given by $\rho_{min}(T)$, as a function of $n_B$ for different temperatures. 
We then consider the relative values of this depletion of the DOS, determined
by $\rho_{min}(T)/\rho_{min}(\infty)$ and cut these 
functions by horizontal lines, lying $4\%$ below the saturation values of 
$\rho_{min}(\infty)$. The crossing points determine the values of $T^*$ 
for any particular $n_{tot}$ and its corresponding value of $n_B$. $T^*$,  
representing a cross-over rather than a phase transition, thus corresponds 
to that temperature where the deviation from the high temperature saturated 
DOS close to the Fermi energy is reduced by an arbitrary but small amount, 
chosen here as $4\%$. In Fig. \ref{TPG} we illustrate the variation of $T^*$ 
as a function of $n_{tot}$ as well as of $(1-n_F)$; the latter being
a measure of the deviation of the Fermion occupation from the half-filled 
band situation and thus of hole doping.

It is illustrative to compare this doping dependence of $T^*$, derived  from 
the dip in the DOS in the normal state, with the doping dependence of the 
mean field critical temperature $T_c^{MF}$ for amplitude fluctuation 
controlled superconductivity. That latter is characterized by the order 
parameters
\begin{equation}
x \; = \; {1 \over N} 
\sum_i \langle c_{i \uparrow}^+ c_{i \downarrow}^+ \rangle, 
\quad \rho \; = \; {1 \over N} \sum_i \langle \rho_i^+ + 
\rho_i^- \rangle
\end{equation}
which refer to the off diagonal elements of the charge operators 
of the electron-pairs and hard-core Bosons respectively. Solving this mean 
field equation problem (for details for such an  analysis the reader is 
referred to an earlier paper\cite{Ranninger-95b}) gives rise to a $T_c^{MF}$ 
which exhibits a doping dependence which is quite similar to that of $T^*$ 
(see Fig. \ref{TPG}), 
with an onset of amplitude fluctuation controlled superconductivity slightly 
below that temperature where electron-pair fluctuations set in. $T_c^{MF}$ 
has of course not the meaning of a transition temperature for the onset of 
superconductivity, which, as we shall see in the next section, is induced by 
phase rather than amplitude fluctuations, except for the limit of low Boson 
concentrations where  $T_c^{MF}$ and $T^*$ smoothly join. 

It is an interesting question to ask how this doping dependence of $T^*$ 
changes when istead of hard-core Bosons one considers normal Bosons. We plot 
in Fig. \ref{TPG} the temperature $T^*_{NB}$, signalling the opening of the 
pseudogap in that case.  It shows a monotonocally decreasing behavior with 
decreasing $n_{tot}$, similar to that found for hard core Bosons but does 
not saturate, as is the case for those latter, when approaching the fully 
symmetric limit $n_{tot}=2$.   

\section{V Spectral properties of the Cooperon propagator}

\begin{figure}
\begin{center}
\includegraphics[width=3.0in,height=3.0in]{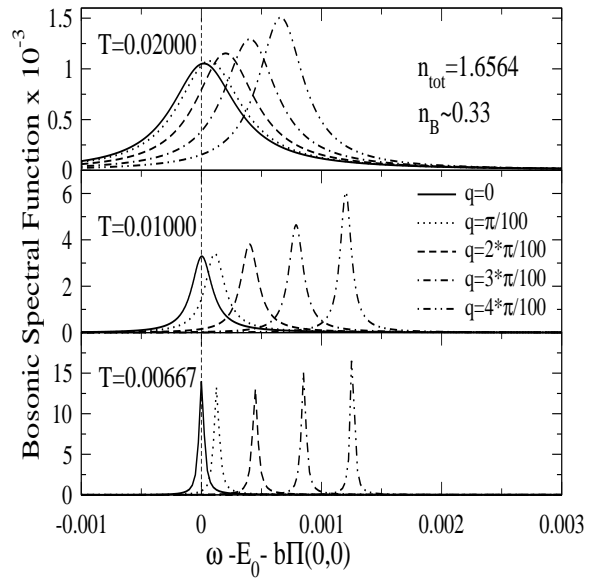}
\end{center} 
\caption{\label{SF1} Temperature evolution of the Boson spectral function 
for the 
low wave vectors for the underdoped regime with $n_{tot}=1.65$.}  
\end{figure}

\begin{figure}
\begin{center}
\includegraphics[width=3.0in,height=3.0in]{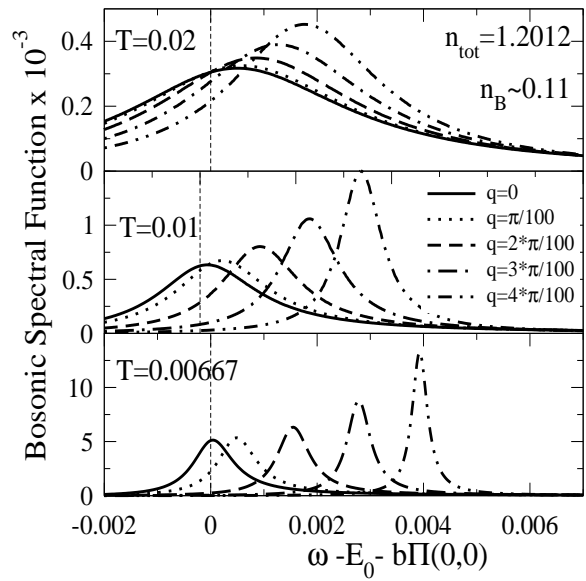}
\end{center}
\caption{\label{SF2} Temperature evolution of the Boson spectral 
function for the 
low wave vectors for the underdoped  regime with $n_{tot}=1.20$.}  
\end{figure}

\begin{figure}
\begin{center}
\includegraphics[width=3.0in,height=3.0in]{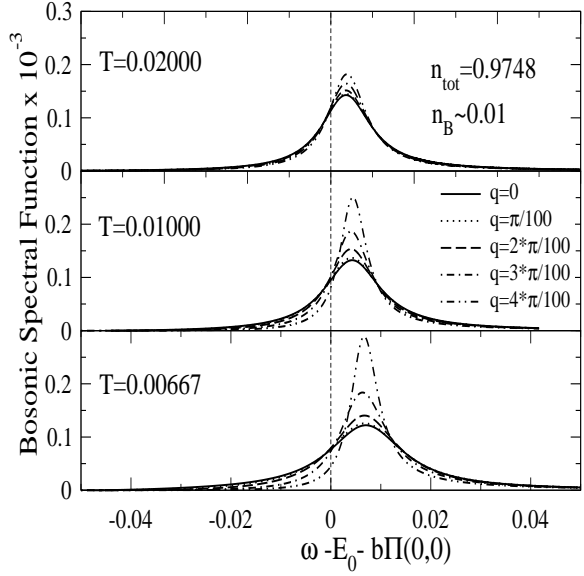}
\end{center} 
\caption{\label{SF3} Temperature evolution of the Boson spectral 
function for the low wave vectors for the optimally/overdoped regime with 
$n_{tot}=0.97$.} 
\end{figure}

\begin{figure}
\begin{center}
\includegraphics[width=3.0in,height=3.0in]{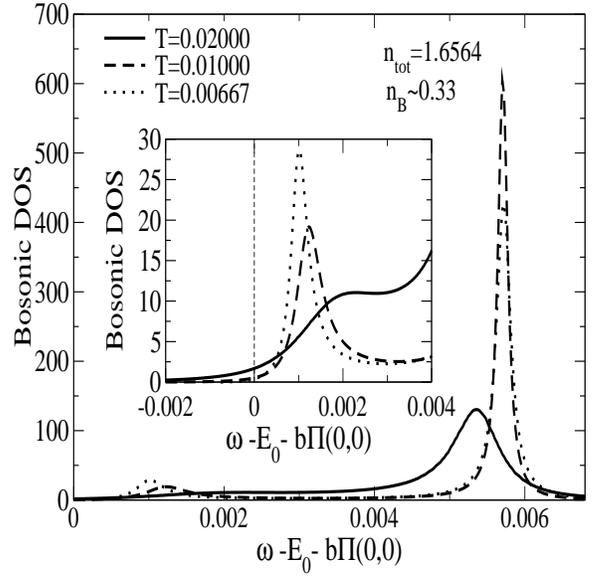}
\end{center} 
\caption{\label{GB1} Temperature evolution of the Bosonic DOS for the 
underdoped regime
with $n_{tot}=1.65$. The insert presents the low energy part of this DOS.} 
\end{figure}

\begin{figure}
\begin{center}
\includegraphics[width=3.0in,height=3.0in]{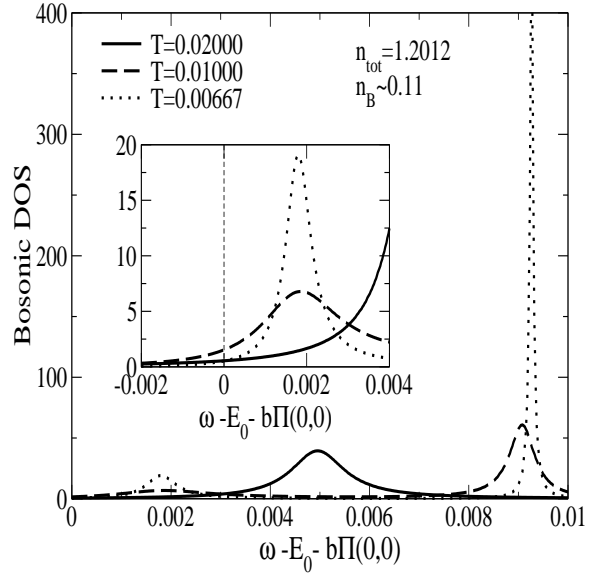}
\end{center} 
\caption{\label{GB2} Temperature evolution of the Bosonic DOS for the 
underdoped regime 
with $n_{tot}=1.20$. The insert presents the low energy part of this DOS.} 
\end{figure}

Let us now examine the features of the normal state which 
act as a precursor of the superconducting phase. As previously 
shown\cite{Devillard-00}, the intrinsically localized bound electron-pairs 
(Bosons), gradually acquire itinerancy as the temperature is lowered below 
$T^*$. We shall here explore this behavior as a function of doping and 
focus on the effect of the hard-core  nature of those Bosons, which has  
been neglected in such previous studies. The resonant electron-pair states in 
the Fermionic subsystem, induced by the exchange with the bound 
electron-pairs of the Bosonic subsystem, are described by the spectral 
properties of the Cooperon propagator
\begin{equation}
C({\bf q},\tau) = \frac{1}{N^2}\sum_{\bf k,k'}
\ll c^+_{\bf {q-k}\uparrow}(\tau) c^+_{\bf {k}\downarrow}(\tau) ; 
c_{\bf {k'}\downarrow}(0)c_{\bf {q-k'}\uparrow}(0) \gg,
\label{G2a}
\end{equation}
which are intimately linked to the spectral properties of the single-particle 
Bose Green's function $K({\bf q}, \omega_m)$ via the relation
\begin{equation}
C({\bf q},\omega_m) = \frac{1}{v^2} \Pi({\bf q}, \omega_m)
+ \frac{1}{v^2} \Pi^2({\bf q}, \omega_m) K({\bf q}, \omega_m).
\label{G2b}
\end{equation}
The thermodynamic and transport properties of our system are given by the 
low lying excitation spectrum of those resonant electron-pairs. Their 
spectral properties are determined by spectral functions of the hard-core 
Bosons, given by the second term in Eq. (\ref{G2b}). In 
Figs. \ref{SF1}-\ref{SF3} 
we illustrate those spectral functions for the long wavelength 
regime together with their evolution with temperature for three representative 
concentrations, which cover the entire doping regime from underdoped 
($n_{tot} \simeq 1.65$) with a high concentration of Bosons to the 
optimal/overdoped ($n_{tot} \simeq 0.97$) with a low concentration of Bosons.
We find that in the underdoped regime the Cooperons are well defined 
propagating modes with a narrow width of the spectral function which, 
moreover, strongly decreases with decreasing temperature   
(see Figs. \ref{SF1}, \ref{SF2}). 
In the optimally/overdoped regime, on the contrary, the spectral functions 
show overdamped mode behavior (see Fig. \ref{SF3}). Tracing the peak position 
of the Boson spectral function as a function of wave-vector ${\bf q}$ permits 
us to determine the mass $m_p$ of those Cooperons. As we approach the dense 
limit of Bosons, $m_p$ increases sensibly, when we compare this mass for 
different values of doping ($n_{tot}$) at a fixed given temperature (see 
Table I). For the low doping regime where the Bosons are well defined 
quasi-particles, their DOS shows an evolution with temperature in 
which the low energy part gets more and more peaked as the temperature is 
lowered and the peak position approaches the value $E_0 +b\Pi(0,0)$, (see 
Fig. \ref{GB1}, \ref{GB2}), as it should according to the Hugenholtz Pines 
theorem\cite{Hugenholtz-59}.

We next turn to the evaluation of the concentration of Cooperons, which, 
together with their mass will enable us to estimate the phase stiffness 
in that system and thus the onset temperature of superconductivity due 
to phase fluctuations, $T_{\phi}$. The density of Cooperons, acting as 
superfluid charge carriers, is contained in the combination of $n_p/m_p$ 
entering the expression for the penetration depth. Alternatively, and
in an approximative fashion, it describes the density of itinerant 
quasi-particles  in the normal state, derivable from the Drude weight in 
the optical conductivity\cite{Tanner-98}. For the BFM scenario investigated 
here, such a Drude component arises from an Aslamazov-Larkin term in the 
conductivity of the itinerant electrons\cite{Devillard-00}, involving the 
Cooperons and is contained in the second term of the Cooperon propagator, 
Eq. (\ref{G2b}). In order to estimate the density $n_p$ of those Cooperons 
which give rise to such a Drude component, we have to attribute it to just  
that contribution of the Cooperon propagator, i.e.,
\begin{equation}
n_p =\frac{1}{N\beta}\sum_{{\bf q}, \omega_m} \frac{1}{v^2} 
\Pi^2({\bf q}, \omega_m) K({\bf q}, \omega_m),
\label{np}
\end{equation}
where the uncorrelated part 
$\frac{1}{v^2} \sum_{{\bf q}, \omega_m} \Pi({\bf q}, \omega_m)$ has been 
subtracted out of the thermal average of the doubly occupied sites, given 
by $<c_{\uparrow}^+c_{\downarrow}^+ c_{\downarrow}c_{\uparrow}>$.
In Table, II we present the mass and and 
concentration of Cooperons for different temperatures and doping rates 
corresponding to the well underdoped ($n_{tot} \simeq 1.65$) and the 
less well underdoped ($n_{tot} \simeq 1.20$) regime, for which the Bosons 
have well defined propagating quasi-particle features.

\vspace{2.0cm}
In order to highlight the effect of the hard-core nature of the Bosons we 
have repeated this study for the Cooperons for the case of normal 
Bosons rather than hard-core Bosons and present the corresponding values for 
the Cooperon  mass and concentration by the numerical values,  given in 
Table I, in the parenthesis.

The above  assessment of the mass and concentration of the Cooperons, 
contributing to the Drude peak in the normal state and ultimately to 
the superfluid current, shows that for hard-core Bosons their concentration 
depends little on doping, while their mass depends on it sensibly. The 
latter rapidly increases, as, upon reducing the hole doping, we enter the 
regime of high Boson concentration where correlation effects become 
increasingly important. As compared to the case of hard-core Bosons, for 
normal Bosons the variation with doping of the concentration of Cooperons 
turns out to be much more important, while for the mass it is less important. 
This  leads, as we shall see below, to significant qualitative differences 
in the respective temperatures determining the onset of phase correlation  
driven superconductivity.
\begin{table}
\caption{\label{table1} Variation with doping $n_{tot}$ of $m_p$ (in 
units of $\frac{1}{D}$) and $n_p$ for a fixed temperature $T= 0.00667$ 
and the estimated resulting $T_{\phi}$. We present in parenthesis the 
corresponding values when the Bosons are treated as normal instead of 
hard-core Bosons.}
\begin{ruledtabular}
\begin{tabular}{llll}
$n_{tot}$ & $m_p$  & $n_p 10^3$ & $T_{\phi}$ \\
\hline
1.651 & 2.27 (1.29)  & 9.60 (14.9) & 0.0042 (0.0116) \\
1.614 & 2.14 (1.24)  & 9.49 (14.5) & 0.0044 (0.0117) \\
1.495 & 1.56 (1.06)  & 9.04 (12.9) & 0.0058 (0.0112) \\
1.376 & 1.14 (0.88)  & 8.44 (11.2) & 0.0074 (0.0127) \\
1.204 & 0.71 (0.62)  & 7.30 (8.71) & 0.0102 (0.0140) \\
1.156 & 0.63 (0.55)  & 6.91 (7.98) & 0.0111 (0.0145) \\
1.105 & 0.45 (0.49)  & 5.88 (7.10) & 0.0124 (0.0145) 
\end{tabular}
\end{ruledtabular}
\end{table}

On the basis of these findings, we now attempt to estimate the critical 
temperature for phase fluctuation controlled superconductivity. For that 
we put $k_B T_{\phi} \simeq \hbar^2 (n_p/m_p) a$, where $n_p$ and $m_p$  
are our estimates for the density $n_s$ and mass $m_s$ of the superfluid 
charge carriers. $a$ denotes a length scale which is of the order of the
coherence length or, alternatively, the inter-plane distance, depending 
on the degree of anisotropy of the system. 

Assuming $a$ in the expression 
for $T_{\phi}$ to be given by the lattice constant, corresponding to a 
layer compound system such as the HTSC, we trace this critical temperature  
as a function of doping ($n_{tot}$ and $1-n_F$) in Fig. \ref{TPG}. We notice a 
crossing of the energy scales related to the phase stiffness and to the 
electron pairing, as we approach the high doping limit, where the density of 
Bosons is small and Cooperons are no longer well defined quasi-particles. 
This phase diagram, Fig. \ref{TPG}, corresponds to that proposed on the 
basis of the phase fluctuation scenario\cite{Emery-95}, but with the 
difference that there the doping dependent quantity of the phase stiffness 
was supposed to be related to the density of superfluid carriers, while, 
according to our present findings, based on the BFM scenario, it should be 
primarily related to the 
mass of the superfluid carriers - estimated as the mass of the Cooperons 
in our case. Upon approaching the optimal/overdoped regime, where 
$T_{\phi}$ crosses $T^*$, the Cooperons loose their good quasi-particle 
features, and the onset of superconductivity is becoming controlled by 
amplitude fluctuations, like in a BCS system. The opposite trend with 
doping of $T_{\phi}$ and $T^*$, observed up to this level of doping, ceases 
accordingly and $T_c$ is constraint to decreases, since being limited from 
above by the decreasing behavior of $T^*$ which controls pair formation. 

In comparison to these features, derived by considering hard-core Bosons, 
we find a noticeably  different  doping dependence 
of the phase fluctuation temperature when these hard-core effects are 
absent and the Bosons are treated as normal Bosons. See the corresponding 
values for $T_{\phi}^{NB}$ in Table I and its graphical representation in 
Fig. \ref{TPG}. $T_{\phi}^{NB}$ decreases only moderately upon approaching 
the low doping regime and, unlike for the case of hard core Bosons, does 
not vanish as the Boson concentration approaches $n_B=1$ for $n_{tot}=2$.

\begin{table}
\caption{\label{table2} Variation of $n_p$, $m_p$ and $n_F$ with temperature 
$T$ for $n_{tot}=1.64$ (top) and $n_{tot}=1.20$ (bottom).}
\begin{ruledtabular}
\begin{tabular}{llll}
$T$ & $n_{p}$ &  $m_p$ & $n_{F}$   \\
\hline
0.00667 & 9.59 & 2.23 $10^{-3}$ & 0.995 \\
0.01000 & 7.86 & 2.63 $10^{-3}$ & 0.994 \\
0.02000 & 6.46 & 5.81 $10^{-3}$ & 0.990 \\
\hline\hline
0.00667 & 7.30 & 0.71 $10^{-3}$ & 0.988 \\
0.01000 & 6.42 & 1.00 $10^{-3}$ & 0.984 \\
0.02000 & 5.56 & 2.91 $10^{-3}$ & 0.971 \\ 
\end{tabular}
\end{ruledtabular}
\end{table}

\begin{figure}
\begin{center}
\includegraphics[width=3.0in,height=3.0in]{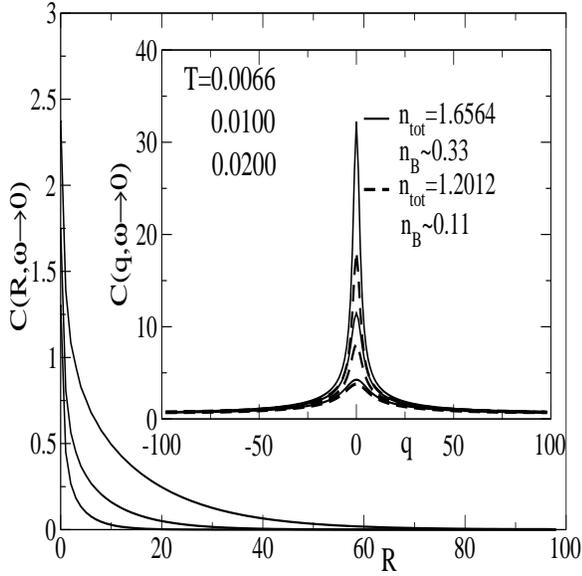}
\end{center} 
\caption{\label{G2} Temperature evolution of the low frequency limit of the 
Cooperon propagator in  the underdoped regime with $n_{tot}=1.65$ 
and $n_{tot}=1.20$ (insert) and its spatial Fourier transform for 
$n_{tot}=1.65$.} 
\end{figure}

In order to illustrate the combined effect of local electron pairing  
and phase correlations of those resonant pair states, we now examine the 
Cooperon propagator, Eq. (\ref{G2b}) in the low frequency limit and 
interpret it in terms of its physically intuitive form, given in eq. (2) 
in section II. Let us for that purpose illustrate this Cooperon 
propagator in Fig. \ref{G2}  as a function of temperature for two cases, 
representing a well underdoped and a less well underdoped  situations, 
showing the tendency with increased hole doping. Fitting the spatial 
dependence of the Cooperon propagator to this phenomenological form, 
permits us to determine the various parameter which characterize it  and 
which we enumerate in Table III. We notice that 
upon increasing  the temperature, the coefficients $C_a$, 
weighting the short range phase-uncorrelated local electron pairing, tend 
to rapidly decrease as we approach $T^*$ in the entire doping 
regime. In the underdoped regime, upon decreasing the temperature and 
approaching $T_{\phi}$, the coefficients $C_{\phi}$, weighting the phase 
correlated electron-pairs, increase rapidly together with the coherence 
length $\xi$, which is typically an order of magnitude 
bigger than the short range electron-pair correlation length $r_0$.
In the optimal/overdoped regime, on the contrary, $C_{\phi}$ hardly changes 
with temperature, while the coherence length follows the similar temperature 
dependence as in the underdoped regime. A systematic change is observed in 
the relative weight of the long range to short range contribution of the 
Cooperon correlation function: $C_{\phi}/C_a$, which, with increased hole 
doping, shows a decreasing behavior when evaluated for some characteristic 
doping dependent temperatures such as $T^*$. In particular, we find 
$C_{\phi}/C_a = 0.71,\, 0.45,\, 0.31$ for $n_{tot}= 1.65,\, 1.20,\, 1.02$ and
corresponding values of $T^* \simeq 0.016,\, 0.0142,\, 0.005$ and 
$T_{\phi}=0.004,\, 0.010,\, 0.014$. It is this relative weight 
increase of amplitude versus phase contributions, as we go from the 
underdoped to the optimal/overdoped regime, which indicates the change-over 
from phase correlation driven superconductivity toward amplitude correlation 
driven superconductivity. A remarkable result is that neither the short nor 
the long range scale depend sensitively on doping. We have some experimental 
indications\cite{Shibauchi-01} from studies in the HTSC that, at least as far 
as $\xi$ is concerned, its doping dependence is very weak.
\begin{table}
\caption{\label{table3}Characteristic parameters of the Cooperon 
propagator, Eq. (2), as a function of temperature $T$ and for three doping 
rates $n_{tot}=1.64$ (top) and $n_{tot}=1.20$ (middle) and $n_{tot}=1.02$ 
(bottom). $\xi$ and $r_0$ are in units of the lattice constant $a$.}
\begin{ruledtabular}
\begin{tabular}{lllll}
$T$ & $C_a$ & $C_{\phi}$ & $r_0$  & $\xi(T)$ \\
\hline
0.00667 & 1.31 & 1.06  & 0.91 & 13.75 \\
0.01000 & 1.13 & 0.61  & 0.72 &  7.59 \\
0.02000 & 0.88 & 0.42  & 0.53 &  3.58 \\
 \hline\hline
0.00667 & 1.23 & 0.66  & 0.83 & 11.76 \\
0.01000 & 1.06 & 0.45  & 0.67 &  6.90 \\
0.02000 & 0.86 & 0.37  & 0.52 &  3.51 \\
\hline\hline
0.00667 & 1.08 & 0.33  & 0.71 &  9.07 \\
0.01000 & 0.98 & 0.32  & 0.62 &  6.18 \\
0.02000 & 0.83 & 0.32  & 0.51 &  3.45 \\
\end{tabular}
\end{ruledtabular}
\end{table}

\section{VI Evidence for pairing correlations in the specific heat}

The onset of the pseudogap, as seen in  numerous experimental 
studies such as ARPES and single particle tunneling, indicate a loss of 
low energy single particle spectral weight. This loss of single particle 
spectral weight ought to be accompanied by a compensating increase of 
spectral weight coming from collective excitations, which, for the 
present precursor scenario,  should predominantly come from pair 
fluctuations. Without having to go to elaborate spectroscopic techniques, 
indications for such Many Body effects are already seen in basic 
thermodynamic quantities such as the specific heat $C_V(T)$ and entropy $S(T)$,
where a hump in $C_V/T$ and a change in slope in $S(T)$ is observed at 
temperatures around $T^*$\cite{Loram-00}. A recent theoretical 
approach\cite{Moca-02} on the basis of a classical pair fluctuation 
scenario\cite{Vilk-97} attributed this hump feature very clearly to the 
contributions coming from pairing correlations, sitting on top of the 
single particle contributions.

\begin{figure}
\begin{center}
\includegraphics[width=3.0in,height=3.0in]{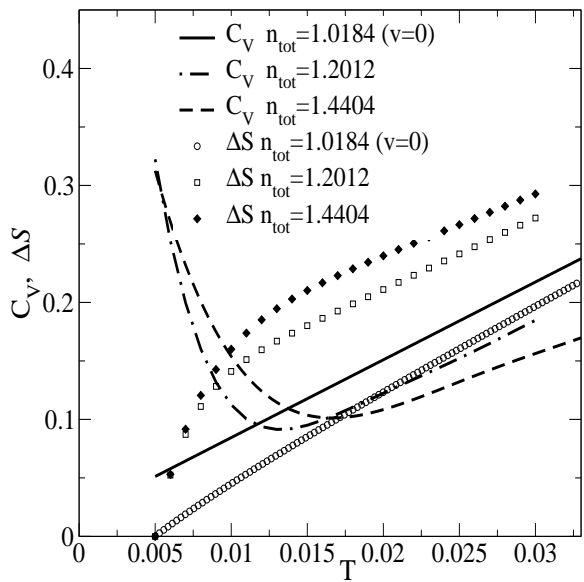}
\end{center} 
\caption{\label{CV} Temperature evolution of the specific heat and 
entropy  in  the 
underdoped regime with $n_{tot}=1.44, 1.20$ and compared with the 
optimal/overdoped regime $n_{tot}=1.02$.}
\end{figure}

We shall in this section present a similar investigation on the basis of the 
two-component precursor scenario adopted here. We evaluate for that purpose 
the inner energy, given by
\begin{eqnarray}
U(T) &=& E_{kin}^F(T) + E_{kin}^B(T) + E_{int}^{BF}(T) \nonumber \\
E_{kin}^F(T) &=& \frac{1}{N}\sum_{\bf k}(\varepsilon_{\bf k}+\mu) 
n_{\bf k}^F(T) 
\nonumber \\
E_{kin}^B(T) &=& \Delta_B n_B(T) \nonumber \\
E_{int}^{BF}(T) &=& v\sum _{i}\langle(\rho^{+}_{i}c_{i\downarrow}
c_{i\uparrow }+\rho^{-}_{i}c_{i\uparrow }^{+}c_{i\downarrow}^{+})  
\rangle_T\nonumber \\
&=& -\frac{2}{N\beta}\sum_{{\bf  q},\omega_m}\Pi({\bf q},\omega_m) 
K({\bf q},\omega_m)
\label{U}
\end{eqnarray} 
and subsequently determine $C_V(T)=\frac{d}{dT}U(T)$ and 
$S(T)=\int_0^T dT'C_V(T')/T'$.
The Fermion distribution function $n_{\bf k}^F(T)$, the number of Bosons 
$n_B(T)$ and the expectation values of the interaction energy have to be 
calculated with respect to the full Hamiltonian. We illustrate in 
Fig. \ref{CV} 
the temperature variation of $C_V(T)$ and $S(T)$ for different total doping 
rates $n_{tot}$, corresponding to the underdoped situation and compare it 
with the non-interacting case ($v=0$) for the case of very small 
concentrations of Bosons. The pseudogap is manifest in the dip-like feature 
of  $C_V(T)$ which occurs at $T^*$, together with a subsequent upturn 
of  $C_V(T)$ upon lowering the temperature which indicates the broad 
hump-like structure above $T_c$. As we decrease $n_{tot}$ upon going from the 
underdoped to the optimal/overdoped regime, the dip moves to lower 
temperatures in 
agreement with a decreasing $T^*$ and eventually disappears upon reaching 
a doping concentration where the number of Bosons tends to zero (given 
approximately by the non-interacting BFM with $v=0$. At the same time the 
linear slope of  $C_V(T)$ at high temperatures increases in correspondence 
with an increase of the DOS at the Fermi energy and saturating at the value 
characterizing the low Boson concentration limit. Concerning the entropy, 
we are able to evaluate its increase, $\Delta S(T)$, starting form a fixed 
lowest temperature and going up to the highest temperature we have been 
considering. The rapid rise in  $\Delta S(T)$ below $T^*$, changing into a 
rather slow rise above $T^*$ signals the existence of local order below 
$T^*$ which gradually disappears when going beyond $T^*$ to higher 
temperatures.

\section{ VII Summary}
 
A doping induced change-over from phase- to amplitude-fluctuation driven 
superconductivity is shown to result in a system with precursor pairing 
within a two-component scenario, involving charge carriers with 
different statistics: Fermions and Bosons, coupled together via a charge 
exchange term. The Fermions describe free electrons while the Bosons (more 
precisely hard-core Bosons) describe localized self-trapped  electron-pairs 
having spin-$\frac{1}{2}$ statistics which  give rise to correlation effects 
in such a system. The opposite variation with doping of $T^*$ and $T_c$ is 
obtained, where the critical temperature is given by the phase 
stiffness of the system. With decreasing the concentration of the 
localized electron-pairs, the energy associated with this phase stiffness 
crosses the pairing energy  $k_B T^*$ in the itinerant electrons subsystem
at a certain characteristic doping level. There, phase fluctuation 
controlled superconductivity changes over into 
amplitude fluctuation controlled superconductivity, giving rise to a phase 
diagram, which, qualitatively, is reminiscent of that proposed for the HTSC  
within the so-called phase fluctuation scenario. The doping dependent 
length scales for short range local electron-pair correlations and long 
range phase correlations are discussed on the basis of the spectral 
properties of the Cooperon propagator, describing the exchange induced 
pairing in the electron subsystem. In the precursor pairing scenario 
studied here, it turns out that it is the degree of itinerancy of the Cooperons
rather than their concentration which controls the doping dependence of the 
phase stiffness. This degree of itinerancy varies from well defined itinerant 
electron-pair states in the limit of high concentration of localized bound 
electron-pairs (low hole doping) to overdamped excitations in the limit of 
low concentration of localized bound electron-pairs (high hole doping). 

The phase diagram, Fig. \ref{TPG}, represented as a function of doping, involves 
changes in the concentration of electrons away from half filling (hole doping) 
for given changes in total concentration of charge carriers, $n_{tot}$. The 
variation of the hole doping (changes in $(1-n_F)$ of the order of $2\%$) in 
this phase diagram is small  compared to  changes in  $n_{tot}$, which is due 
to the $1D$ situation we have been considering here. This result is less 
surprising when we consider a $2D$ system with an anisotropic 
charge exchange coupling between the bound electron-pairs and the bare 
itinerant electrons. One would then obtain a corresponding anisotropic 
pseudogap, similar to what is observed in the HTSC. 
Hole doping would now affect roughly equally all the regions near the 
$2 D$ Fermi surface, and thus attributing only a very small fraction of 
the doped holes to the actual regions in the Brillouin zone where the 
pseudogaps are formed, i.e., around the $M$ points and along lines parallel 
to $[0,0] - [0,\pi]$ and equivalents.

One of the outstanding problems to be solved within such a precursor scenario 
for superconductivity is to understand how the transition to the 
superconducting state occurs. This involves a competition between 
Cooper pairing and real space pairing and necessitates a generalization 
of the present system of Green's functions by including the anomalous 
Green's functions in order to treat superconducting fluctuations. This 
problem will be the issue of future studies. 
 
\section{Acknowledgment}
The authors are extremely grateful to B. K. Chakraverty, T. Domanski and 
G. Jackeli for innumerable discussion on this subject as well as with help 
in setting up the diagramatic procedure for hard-core Bosons and assistance
with numerical procedures.

\bibliography{biblio1}

\end{document}